\def\gtorder{\mathrel{\raise.3ex\hbox{$>$}\mkern-14mu
             \lower0.6ex\hbox{$\sim$}}}
\def\ltorder{\mathrel{\raise.3ex\hbox{$<$}\mkern-14mu
             \lower0.6ex\hbox{$\sim$}}}
\def\ltsima{$\; \buildrel < \over \sim \;$}
\def\simlt{\lower.5ex\hbox{\ltsima}}
\def\gtsima{$\; \buildrel > \over \sim \;$}
\def\simgt{\lower.5ex\hbox{\gtsima}}
\def\hst{{\it HST }}
\def\Ha{H$\alpha$ }
\def\Hb{H$\beta$ }
\documentstyle[12pt,aaspp4]{article}

\begin{document}

\title{The Ultraviolet Spectra of LINERs: A Comparative Study\footnote
{Based on observations with the {\it Hubble Space Telescope} which is operated
   by AURA, Inc., under NASA contract NAS 5-26555}
}

   \author{Dan Maoz\altaffilmark{2}, Anuradha Koratkar\altaffilmark{3},
 Joseph C. Shields\altaffilmark{4},}
 \author  { Luis C. Ho\altaffilmark{5},
Alexei V. Filippenko\altaffilmark{6}, \& Amiel Sternberg\altaffilmark{2,6}}

 \altaffiltext{2}{School of Physics \& Astronomy and Wise Observatory,
    Tel-Aviv University, Tel-Aviv 69978, Israel. dani@wise.tau.ac.il, amiel@wise.tau.ac.il}

 \altaffiltext{3}{Space Telescope Science Institute, 3700 San Martin Dr., Baltimore, MD 21218.
    koratkar@stsci.edu}

  \altaffiltext{4}{Physics and Astronomy Department, Ohio University, Athens, OH 45701.
    shields@helios.phy.ohiou.edu}

   \altaffiltext{5}{Harvard-Smithsonian Center for Astrophysics, 
    60 Garden Street, Cambridge, MA 02138. lho@coyote.harvard.edu}

    \altaffiltext{6}{Department of Astronomy, University of California, Berkeley, CA 94720-3411.
   alex@astro.berkeley.edu}

\begin{abstract}
Imaging studies have shown that $\sim 25\%$ of LINER galaxies
display a compact nuclear UV source.
As part of a program to study the nature of LINERs and their
connection to the active galaxy phenomenon,
we compare the \hst ultraviolet (1150--3200~\AA) spectra  of seven 
such UV-bright LINERs.
Data for three of the galaxies (NGC 404, NGC 4569, and NGC 5055)
are presented for the first time, while data for four others
(M81, NGC 4594, NGC 4579, and NGC 6500) have been recently
published. The spectra of NGC 404, NGC 4569, and NGC 5055 show clear
absorption-line
signatures of massive stars, indicating a stellar origin for the UV 
continuum.
Similar features are probably present in NGC 6500.
The same stellar signatures {\it may} be present but undetectable in NGC 4594,
due to the low signal-to-noise ratio of the spectrum, and in M81 and NGC 4579, due
to superposed strong, broad emission lines.
The compact
central UV continuum source that is observed in these galaxies is
a nuclear star cluster rather than a low-luminosity active galactic
nucleus (AGN), at least in some cases. 
Except for the two LINERs with 
broad emission lines (M81 and NGC 4579), the LINERs have 
weak or no detectable UV emission lines. The UV emission-line
spectrum strength shows no relation to the UV continuum strength.
Furthermore, at least four of the LINERs suffer from an ionizing photon
deficit, in the sense that the ionizing photon flux inferred from the
observed far-UV continuum is insufficient to drive the optical H~I 
recombination lines.
Examination of the nuclear X-ray flux of each galaxy shows a high
X-ray/UV ratio  in the four ``UV-photon starved''
LINERs. In these four objects, a separate component, emitting predominantly in the
extreme-UV, is the likely ionizing agent, and is
perhaps unrelated to the observed nuclear UV emission.
Future observations can determine whether the UV continuum in
LINERs is always dominated by a starburst or, alternatively,
that there are two types of UV-bright LINERs: starburst-dominated
and AGN-dominated.

\end{abstract}


\section{Introduction}

Low-ionization
nuclear emission-line regions (LINERs) are detected in the
nuclei of a large fraction of all bright nearby galaxies
(Ho, Filippenko, \& Sargent 1997a).
Since their definition as a class by Heckman (1980), they have elicited debate
as to their nature and relation, if any, to active galactic
nuclei (AGNs). On the one hand, the luminosities of most LINERs
are unimpressive compared to ``classical'' AGNs, and can easily be produced 
by processes other than accretion onto
massive black holes. Indeed, LINER-like spectra are sometimes
seen to arise in ``non-nuclear'' environments, such as cooling flows.
On the other hand, a variety of observables
point to similarities and continuities between AGNs
and at least some LINERs (see the contributions in Eracleous et al. 1996, for reviews). 
If LINERs represent the low-luminosity end of the AGN phenomenon,
then they are the nearest and most common examples, and their 
study is germane to understanding AGN demographics, quasar evolution,
dormant black holes in quiescent galaxies, the X-ray background,
and the connection between AGN-like and starburst-like activity.

The ultraviolet (UV) sensitivity and angular resolution of the {\it Hubble Space Telescope}
(\hst) is providing new clues toward understanding LINERs.
Maoz et al. (1996a) carried out a UV (2300 \AA) imaging survey of the 
central regions of 110 nearby galaxies with the Faint Object Camera (FOC)
on {\it HST}. As reported in Maoz et al. (1995), five among the 25 LINERs
in their sample revealed nuclear UV sources, in most cases unresolved,
implying physical sizes $\ltorder 2$ pc. Maoz et al. (1995) argued
that the UV sources in these ``UV-bright'' LINERs could be the
extension of the ionizing continuum, which is rarely seen in LINERs
at optical wavelengths due to the strong background from the normal
bulge population. The compactness of the sources suggested that they could be 
nonstellar in nature, although compact star clusters of such
luminosity were also possible. A similar fraction of UV-bright
nuclei was found in a 2200 \AA\ imaging survey of 20 LINER and
low-luminosity Seyfert 2 galaxies carried out with the WFPC2 camera on
\hst by
Barth et al. (1996a; 1998). A compact, isolated, nuclear UV source
has also been found in FOC images of the LINER NGC 4594 at $\sim 3400$ \AA\ 
 by Crane et al. (1993), and in  WFPC2 images of the LINER M81 in a broad
(1100 \AA\ to 2100 \AA) bandpass by Devereux, Ford, \& Jacoby (1997).
  The UV-bright LINERs were obvious
targets for follow-up spectroscopy with \hst. Faint Object Spectrograph
(FOS) observations have
been analyzed for the LINERs M81 (Ho, Filippenko, \& Sargent 1996), NGC 4579
(Barth et al. 1996b), NGC 6500 (Barth et al. 1997), and NGC 4594
(Nicholson et al. 1998). Among these four, NGC 4579 was 
part of the Maoz et al. (1995) FOC survey and NGC 6500 was in the 
Barth et al. (1996a, 1998) WFPC2
survey.

The spectra of M81 and NGC 4579, which in the optical range have
weak broad wings in the \Ha line (Filippenko \& Sargent 1985; 1988;
Ho et al. 1997b), are reminiscent of AGNs  in the UV, with strong, broad emission
lines superposed on a featureless
continuum. Ho et al. (1996) and Barth et al. (1996b) concluded that
these two LINERs are most probably AGNs. NGC~6500 and NGC 4594, on the other hand,
have only weak and narrow emission lines on top of a UV continuum.
Based on the
resolved appearance of the UV source in the \hst WFPC2 F218W image,
and the tentative detection of optical Wolf-Rayet features, Barth et al. (1997)
concluded that the UV emission in NGC~6500 is likely dominated by 
massive stars, though contribution from a scattered AGN
component could not be excluded. For NGC 4594, Nicholson et al. (1998) favor
an AGN interpretation.

In this paper, we present FOS UV spectra of three additional
LINERs (NGC 404, NGC 4569, and NGC 5055), and compare the
properties of all seven UV-bright LINERs observed spectroscopically
with \hst to date.

\section{Observations}

The three LINERs with new data were observed with a similar
observational setup to that used for the four previously published LINERs.
The FOS $0.86''$ diameter circular aperture was centered on the nucleus using
a three-stage peakup, and three grating/detector combinations were used: 
G130H/FOS-BLUE (1152--1608 \AA, 1.0 \AA\ diode$^{-1}$),
G190H/FOS-RED (1600--2280 \AA, 1.5 \AA\ diode$^{-1}$),
and G270H/FOS-RED (2222--3277 \AA, 1.0 \AA\ diode$^{-1}$).
The spectral resolution is about half a diode.
Table 1 lists the main observational parameters for the
seven objects. 
Among the previously studied LINERs, M81 and NGC 4579 were
 observed with the $0.3''$-diameter and $0.43''$-diameter
apertures, respectively. This should
make little difference compared to the $0.86''$ apertures
used for the other objects, since Maoz et al. (1995) demonstrate
that the main central UV source in NGC 4579 has FWHM $< 0.11''$,
and Devereux et al. (1997) find the UV source in M81 has FWHM
$< 0.1''$.
Similarly, the UV images of most of the objects (Maoz et al. 1995;
Barth et al. 1996a, 1998; Devereux et al. 1997) show that all the UV flux from the 
central compact UV source would be included in the FOS aperture,
and light from neighboring sources excluded. This statement
holds with less confidence for NGC 4594, where UV imaging data
exist only in the near UV (3400 \AA; Crane et al. 1993),
but the nucleus appears unresolved and isolated
at that wavelength.

The spectra of the three LINERs with new data were reduced,
as in the case of the previously studied objects, by the 
standard \hst calibration pipeline, which includes flat-fielding,
subtraction of the particle induced background, and flux
and wavelength calibration. We reiterate the conclusions
of Barth et al. (1997) and Nicholson et al. (1998) that in the two faintest
objects, NGC 6500 and NGC 4594, the correction for scattered
light toward short wavelengths in the G130H grating is only approximate.
In NGC 4594, the G130H flux is dominated by scattered light,
so the continuum shape measured in that spectral region is unreliable.

Table 1 also gives a comparison of the observed $\sim 2200$ \AA\ flux density 
measured from the FOS spectra with the flux from \hst
broad-band measurements at that wavelength.
Such a comparison is relevant
for appraising the accuracy of the spectroscopic measurements
and looking for evidence of variability. For M81, we have used the
FOS spectra and the WFPC2 F160BW throughput curve
to derive the 1500 \AA\ continuum flux that would produce the count
rate observed in the WFPC2 image  of Devereux et al. (1997).
 Note that the level
of agreement between the various measurements of the same object
is mixed. The uncertainty of the FOC F220W measurements of individual
sources is $\sim 20\%$, due to flat-fielding and background
uncertainties (Meurer et al. 1995). 
The uncertainty is larger for sources which Maoz
et al. (1995) could not fit properly with the pre-COSTAR
 \hst point-spread
function, because of additional diffuse and compact sources
near the main central UV source (e.g., in NGC 404 
and NGC 5055). For M81, the variable sensitivity of 
WFPC2 in the far-UV and the unknown spectrum of the object in the G190H 
range both contribute to the uncertainty. 
From an analysis of the FOS target acquisition records, we deduce 
that NGC 404 and M81 were found at the edges of the peak-up scans,
possibly compromising the photometric accuracy of their observations.
In view of all this, the FOC, WFPC2, and FOS measurements
of a given object in Table 1 are consistent, except for
the case of NGC 4579. 

As discussed by Barth et al. (1996b), in NGC 4579
the FOS 2200 \AA\ flux is less than 1/3 that implied by the FOC
measurement made 19 months earlier. Barth et al. (1996b)
interpreted the difference as due to variability of a nonstellar
continuum source. However, there is a chance
that the low FOS flux was the result of a miscentered aperture.
We note that
there is a second, slightly extended UV source $0.58''$ east
of the main nuclear source, which has about 1/6 of
its flux  (see Maoz et al. 1995). The FOS $0.43''$ aperture
may have been accidentally centered between this source (which might be relatively
brighter in the G270H grating's wavelength range, in which the peakup procedure
was done) and the central source,
bringing the edge of the aperture, say, $0.1''$ from the central source.
Half of the flux in the observed spectrum would then be from the
secondary source and half from the fraction of the point-spread function
 of the central source
within the aperture. We have re-analyzed the \hst records of the
target acquisition for this observation, and they indicate that
the FOS aperture was properly centered on the brighter of two
sources, if two were present. Furthermore,
a broad visual-band (F547M) WFPC2 exposure of the galaxy (to be
described elsewhere) shows a single, unresolved nucleus, with no
trace of an additional source to the east. This supports the
variability interpretation.
It would be valuable, however, to confirm such
variability in this and other LINERs. 

We also list in Table 1 the Galactic reddening, $E(B-V)$, that we have
adopted for each object, based on the H I column density $N_H$ from
 Murphy et al. (1996) if available,
with the conversion $E(B-V)=N_H/5.8\times 10^{21}$ cm$^{-2}$,
or using Burstein \& Heiles (1984). For NGC 6500, we have followed Barth
et al. (1996b) and used the Burstein \& Heiles (1984) value, despite
the availability of a measurement by Murphy et al. (1996).
Correcting the NGC~6500 spectrum using the Murphy et al. (1996) value  leads
to a distinct bump in the continuum at 2200 \AA. 

Figures 1--4 show sections of the spectra of the seven LINERs.
We analyze and compare them below.

\section{Analysis}
\subsection{Far-UV Spectral Signatures of Massive Stars}

Figure 1 shows most of the G130H spectrum of all
seven LINERs. (The extreme red and blue ends are excluded.)
 The objects are ordered with the two broad-lined
LINERs  (M81 and NGC 4579) on top, and the other LINERs in order
of decreasing $f_{\lambda}$. Figure 2 shows in more detail
the same spectral
region for  NGC 4569, NGC 404,
NGC 5055, and NGC 6500. Here and in the following figures
all the spectra have been shifted to the object's rest frame
(according to the velocities listed in Table 1), and binned
in 1 or 2 \AA\ bins to improve the signal-to-noise ratio (S/N). 
Overlayed on each spectrum
(thin line) is a 
scaled, normalized version of the \hst Goddard High Resolution Spectrograph
 spectrum of the 
``B'' clump in NGC 1741, a starburst galaxy, which
is described by Conti, Leitherer \& Vacca (1996).
The broad, blueshifted absorption
 profiles of C IV $\lambda1549$ and Si IV $\lambda1400$
 (and also N V $\lambda 1240$) in NGC 1741-B
are the signatures of winds produced by massive
stars (see Leitherer, Robert, \& Heckman 1995). 
The other, narrower, 
absorption lines are of both photospheric and interstellar-medium (ISM) origin.
The characteristic starburst features seen in the spectrum of
 NGC 1741-B are also observed in star-forming regions in 30 Doradus
(Vacca et al. 1995) and NGC 4214 (Leitherer et al. 1996), in 
star-forming galaxies at redshifts $z\approx 2-3$ (Steidel et al. 1996;
 Lowenthal et al. 1997),
and in the $z=3.8$ starburst radio galaxy 4C 41.17 (Dey et al. 1997).

The FOS spectrum of NGC 4569, the brightest (and highest S/N)
LINER in our sample, is virtually identical to that of NGC 1741-B.
Most of the features in the two spectra match one-to-one.
NGC 4569 has all the main interstellar absorption lines seen in starburst 
spectra, as well as the broad stellar wind features. Also
detected are several narrow absorption lines which
cannot be interstellar (they are not 
resonance lines) and constitute further evidence for the photospheres 
of hot stars: Si~III $\lambda\lambda 1294/1297$, 
Si~III $\lambda1417$, C III~$\lambda 1427$, S V $\lambda 1502$, and
N IV $\lambda 1720$ (Heckman \& Leitherer 1997).
The correspondence between the two spectra is such that 
a dilution of the stellar features by a nonstellar (i.e., featureless)
 continuum producing more than 20\% of the UV continuum would be
readily apparent.
The only significant difference between the spectra of the starburst
clump in NGC 1741 and the nucleus of NGC~4569 is the presence of
relatively strong 
C I absorption lines (at rest wavelengths 1277~\AA, 1280~\AA, 1329~\AA,
1561~\AA, and 1657~\AA) in NGC 4569.\footnote{
The C I lines are generally not seen with such strength
(up to 3.5 \AA\ equivalent width) in starburst spectra. 
Interestingly, the same
C I lines seem to be present in some of the other LINERs as well, so they
may be a signature of a particular sort of environment. We discuss the 
interstellar lines in more detail in an appendix.} All the absorption lines 
are at the slight blueshift ($-223$ km s$^{-1}$) of the galaxy,
and hence are not produced in the Milky Way's ISM.
Table 2 lists the main absorption lines detected.
There is no doubt that the UV emission 
in this object, even though its source is highly compact ($\ltorder 2$ pc; Maoz et
al. 1995), is dominated by a cluster of massive stars. Keel (1996) reached
a similar conclusion regarding the optical continuum emission, based
on the presence of narrow Balmer-line absorption, characteristic of
A-type supergiants. The luminosity of a star cluster is dominated by
the most massive stars. Using spectral synthesis models (see \S 3.3 )
and assuming a distance of 9 Mpc (Rood \& Williams 1993) to this galaxy,
we estimate that 250-600 O-type stars (depending on the presence of very massive
early-O-type stars) are sufficient to produce
the observed luminosity. (In our models, O-type stars are defined as
stars having temperatures $T \ge 30,900 K$.) 
As discussed in more detail in \S 3.2, below, a plausible
amount of dust extinction can raise this number by factors of up to
several tens.

Proceeding to the far-UV spectrum of NGC 404 (Fig. 2), we note that 
the broad, blueshifted C~IV absorption is shallower,
but definitely detected.  Si IV absorption
is not seen,
except in the narrow components. According to the synthesis by
 Leitherer et al. (1995), such spectra characterize 
a starburst that is either very young ($<3$ Myr) or $>5$ Myr old. 
The denser winds of O-type supergiants, which are present only
during a limited time, are required in order to produce the broad blueshifted Si IV
absorption. The C~IV absorption, on the other hand, remains as long as 
there are main-sequence O-type stars. The relative shallowness of the C~IV
profile can also be reproduced well by diluting the NGC 1741-B starburst
 spectrum by a featureless (e.g. nonstellar) 
spectrum that is constant in $f_{\lambda}$, and contributes
$\sim 60$\% of the total flux. The strongest interstellar absorption
lines that appear in NGC 4569, including the C I lines, are seen in NGC 404
as well. Due to the small blueshift of this galaxy ($-36$ km s$^{-1}$),
we cannot say whether these lines arise in the Milky Way or in NGC 404
based on their wavelengths alone. However, \hst studies of Milky Way
absorption along the lines of sight to quasars (Savage et al. 1993)
shows that the equivalent widths of Milky Way absorption in
these transitions is several times smaller than the equivalent widths
listed for the LINERs in Table 2. For comparison, we include in Table 2
the range of Galactic equivalent widths measured by Savage et al. (1993).
The velocity width and blueshift of the C~IV profile
clearly point to a stellar origin for this line.
 (The ``emission'' between 1295~\AA\ and 1310~\AA\ in NGC 404 
is an artifact due to a noisy diode
 that masks the O I and S II absorptions that 
are probably present at those wavelengths.) It thus appears that 
the UV continuum source in this
LINER is a star cluster that is of different age than the one in
NGC 4569, or perhaps a 
cluster of similar age whose emission is diluted by a nonstellar continuum.
Assuming a distance of 2 Mpc (Tully 1988) to NGC 404, its UV luminosity
is 100 times lower than that of NGC 4569, implying just two to six 
O-type stars can produce the observed luminosity. Again, the actual number
is likely higher after a reasonable extinction correction.

Continuing to the G130H spectrum of NGC 5055 (Fig. 2), 
the S/N degrades, but 
the blueshifted broad C IV and Si IV 
absorptions seen in NGC 4569 are definitely present.
Their minima are at the galaxy's redshift  ($497$ km s$^{-1}$).
The properly redshifted minimum, the width, and the blue
asymmetry of the C~IV profile all argue against the
possibility that we are merely seeing a blend of strong
Milky Way or host galaxy ISM absorption lines. 
The absorption lines may be slightly broader than in NGC 4569.
From the G270H spectrum,
which has higher S/N, we also find that the dominant ISM
absorptions are in the host galaxy, not the Milky Way.
The stellar nature of this UV source is not  surprising,
since Maoz et al. (1995) already noted that it is
marginally resolved at 2200 \AA, with FWHM$\approx 0.2''$ ($\approx 6$ pc).
NGC 4569 and NGC 5055 are remarkably similar
over the entire UV range.
Overall, the spectrum of NGC~5055 resembles a lower S/N version
of that of NGC~4569. At a distance of 6 Mpc (Phillips 1993), NGC 5055 is $\sim 20$
times less luminous than NGC 4569.

In the spectrum of NGC 6500 (Fig. 2), 
the blueshifted broad C IV absorption is
possibly recognizable although, as noted by Barth et al. (1997), its
significance is arguable when the spectrum is viewed individually.
 The comparison to the spectra of the
other LINERs and the starburst spectrum,
 combined with the obvious degradation in S/N in
this fainter object, suggest that the absorption may, in fact, be present.
The spectrum is too noisy and dominated by scattered light
to reach any conclusion regarding the Si IV absorption. As noted
above, Barth et al. (1997) already concluded, based on the
extended ($0.5''\approx 100$ pc) appearance of the UV source in a WFPC2
image, that this source is probably stellar in nature.

Finally, the NGC 4594 G130H data (Fig. 1) are too noisy and dominated
by scattered light to reach any conclusion, except that
they could be consistent with the same type of spectrum.
There are certainly no strong {\it emission} lines in this part 
of the spectrum of NGC~4594.

We conclude that, in the five LINERs without  
emission lines in the G130H range, the UV continuum is, with varying degrees
of certainty, produced by massive, young stars. 
Could the same be true of the two broad-lined LINERs
in our sample, M81 and NGC 4579? As already noted by Ho et al.
(1996) for M81, one cannot answer this question based on these
UV data alone. Figure 1 shows that if the continuum spectrum
were of the same starburst type as in the other LINERs, we 
would not know it because
the broad emission lines are coincident with the broad
absorptions. 
A case in point
is the recent study by Heckman et al. (1997) of the Seyfert 2
galaxy Mrk 477. The relatively narrow emission lines are
superposed on the broad stellar-wind signatures, but in Si IV and N V
enough of the stellar absorptions are visible near the blue wings
of the emission lines to reveal the starburst nature of the UV continuum
emission.  In M81 and NGC 4579, however, the
emission lines are too broad to see the stellar absorptions, if
they are there.

One may alternatively hope to detect the narrow interstellar 
absorption lines that also typify starburst spectra (or to
put upper limits on the presence of such lines). We see, however,
that even in NGC 4579, which has a higher S/N G130H spectrum than
M81, the interstellar lines would not be detected even if they were
present. Indeed,
in the regions between the broad emission lines, the scaled starburst
spectrum appears more ``featureless'' than the continuum of NGC 4579. 
A S/N of about 20 in the continuum, as in NGC 4569, rather than
about 5, as is the case in NGC 4579, would allow the detection of
interstellar lines in the spectrum.
The single argument that the UV continuum source in NGC 4579 is 
necessarily nonstellar is its  possible factor $\sim 3$ variability reported in  
Barth et al. (1996b).

The compact UV continuum sources in three or four of the seven LINERs
are therefore clusters with massive stars. Barring perhaps
NGC 4579 (assuming its UV variability is real), the UV
continuum in {\it all} the LINERs could be stellar in origin.
Note that this result does not bear directly on the question
of whether or not there is also a nonstellar, quasar-like
object in the nucleus. A microquasar could still
be present, and dominate the emission at wavelengths other
than the UV. Indeed, we will show below that additional
continuum energy, beyond what is implied by the stellar UV
continuum, is required by the emission-line energy budget
in some of these LINERs. Furthermore, it is becoming
increasingly appreciated that circumnuclear
starbursts, albeit on physical scales larger than those
considered here, can contribute 
significantly to the UV and optical brightness of some
AGNs (e.g., IC 5135 -- Shields \& Filippenko 1990; Heckman et al. 
1998; NGC 1068 -- Thatte et al. 1997; NGC 7469 --
Genzel et al. 1995; Mrk 477 -- Heckman et al. 1997; 
some Seyfert 2s -- Heckman 1998).

\subsection{UV Spectral Energy Distribution and Extinction}

Figure 3 shows the UV spectral energy distributions (SEDs)
 of the seven LINERs in $\log f_{\lambda}$ vs. $\log \lambda$,
and in the
same order as in Figure 1.
The spectra have been corrected for Galactic extinction
using the extinction curve of Cardelli, Clayton, \& Mathis (1989), assuming
$A_V=3.1E(B-V)$ and the $E(B-V)$ values listed in Table 1.
The spectra have been vertically shifted, as indicated
on the right side, for clarity. The G130H spectrum of NGC 4594
is not plotted because it is dominated by scattered light
and hence unreliable.
The straight lines 
drawn through every spectrum at $\log\lambda=3.36$ (2300 \AA)
are power laws $f_{\lambda}\propto \lambda^{\beta}$, with
$\beta=0,-1, -2$. The figure shows that the seven LINERs
have UV SEDs that are similar. After the
correction for Galactic reddening, most can be 
roughly described shortward of 2300 \AA\ by power laws 
with $\beta \approx -0.5$ to $0.5$. Table 3 lists $\beta$
for the seven galaxies.
NGC 4579 is again somewhat
exceptional in that its slope changes shortward of 1900~\AA\ 
from $\beta\approx 0$ to $\beta\approx -1.5$ (Barth et al. 1996b).
Typical Seyfert~1 AGNs have power-law continua 
with $\beta$ of $-1$ to $-1.5$ in the UV.
 The intrinsic spectra of unreddened
massive stars, and hence of synthesized young starbursts, rise
even more steeply to the far UV, with
$\beta \approx -2$ to $-2.5$ (Leitherer \& Heckman
1995). Many observed starbursts also have similar slopes
(Kinney et al. 1996), though not quite as steep as $\beta=-2.5$,
 even when their Balmer decrements indicate no reddening. 
An empirical starburst template spectrum derived by 
Calzetti (1997a) has $\beta=-2.1$.
If the similarity of the LINER SEDs
is the result of reddening, it is puzzling
that the reddening is generally tuned to produce a flat
spectrum in $f_{\lambda}$, i.e., a change in slope
between the intrinsic and the reddened spectrum,
$\Delta\beta\approx 2$.
The extinction law must be
different from the Galactic one, in order not to introduce a strong
2200~\AA\ feature. A similar conclusion has been found 
to hold generally for starbursts by Calzetti, Kinney, \& 
Storchi-Bergmann (1994, 1996).

UV extinction curves vary greatly depending on the environment
and the geometry of sources, dust, and gas.
To gauge the amount of extinction suffered by the UV sources
in the LINERs, we first estimate the extinction from
the slope $\beta$ of the UV continuum. A change in slope
$\Delta\beta$ 
corresponds to a reddening in magnitudes
$$
A({\rm 1300~\AA})-A({\rm 2300~\AA})=2.5 \Delta\beta \log (2300/1300).
$$
The translation of reddening to UV extinction depends
on the assumed extinction curve. Following Pettini et al. (1998),
we take two extreme possibilities of extinction curves
devoid of a 2200 \AA\ bump. One is the steep extinction curve
found for stars in the Small Magellanic Cloud (SMC) by 
Bouchet et al. (1985), as normalized by Pei (1992). 
The other is the ``attenuation
curve'' empirically derived by Calzetti et al. (1994)
from the spectra of star forming galaxies, as normalized in
Calzetti (1997b). The latter is ``greyer''
than the former, so will produce more UV extinction for
a given amount of reddening. For the SMC curve, $\Delta\beta=2$
corresponds to $A_V=0.4$ mag and $A_{1300}=2.4$ mag.
For the Calzetti et al. (1994) curve, one obtains
$A_V=1.9$ mag and $A_{1300}=4.8$ mag. A plausible range
for the attenuation factor at 1300 \AA\ is therefore 10 to 80,
at least for the ``clearly-stellar'' LINERs, where we can be
fairly confident about the intrinsic, unreddened spectral slope.

An independent estimate of the UV extinction in these objects
can be obtained
from their Balmer decrements. The decrements, as measured by Ho 
et al. (1997a), are listed
in Table 3. The \Ha/\Hb ratio in NGC 4579, NGC 5055, and NGC 4569
is in the range 5 to 5.4. Assuming the steep SMC curve, this 
gives a rather implausible 
$A_{1300}=9.4$ mag, while the Calzetti et al. curve
gives $A_{1300}=4.4$ mag. In the other four 
galaxies, the Balmer decrement is in the range 3.4 to 3.7,
implying, after accounting for the Galactic reddening, 
$A_{1300}=2.3$ mag from the SMC curve
and $A_{1300}=0.7$ mag using Calzetti et al.'s curve.

The two reddening estimators, based on UV slope and on
Balmer decrement, are obviously inconsistent, if only
because the similar UV slopes 
of most of the objects suggest similar UV extinction, 
while the Balmer decrements indicate
that some are highly reddened and some are not. A possible
solution is that the continuum sources and the line-emitting
gas have different dust distribution geometries, with either
the gas or the continuum sources undergoing more extinction,
depending on the reddening curve assumed.
An alternative solution is that different reddening curves operate
in the different objects. The two reddening estimators are 
broadly consistent if the three objects with high Balmer decrements
are reddened by the Calzetti et al. curve, while the four
objects with low Balmer decrements undergo SMC-like reddening.

\subsection{UV Emission Lines and the Ionizing Photon Budget}

Except for M81 and NGC 4579, the LINERs have few or no detectable
UV emission lines in the G130H range\footnote{Figure 1 of Barth et al. 
(1997) shows that Ly $\alpha$ is resolved
from geocoronal emission in the spectrum of NGC 6500. Also, the
extremely weak emission line at 1240~\AA, if real, 
may be Mg~II $\lambda\lambda$1239.92, 1240.40, 
rather than N~V (Wallerstein 1998).}. Some of them 
do display emission lines in the mid-UV range,
as illustrated in Figure 4.
The 1620 -- 2500 \AA\ range (i.e., G190H and part of
the G270H range) is plotted, with the objects
ordered from top to bottom with decreasing equivalent
width of the main lines. Note how the ratio between
the two strongest lines, 
C~III]~$\lambda$1909 and C~II]~$\lambda$2326, is
approximately constant as long as they are detected,
while there is a variation by over two orders of magnitude in
the equivalent width of these lines among the objects. The
lines are barely detected in NGC~404, and undetected
in NGC~5055 and NGC~4569, despite the high S/N in these
three objects. (The emission at 2125 \AA\ in NGC 6500 is
an artifact -- see Barth et al. 1997.)  
The UV continuum and the flux in these UV lines
are clearly not correlated.
This suggests that whatever is generating
the UV continuum is not the main driver of the emission lines.

Table 3 lists the mid-UV emission-line fluxes in the seven
objects. The fluxes are from a direct integration above
a straight-line continuum, and for the broad-lined objects
include the total (broad- and narrow-line) flux. The 
Mg II $\lambda$2800 emission line is not listed,
as it is clearly affected to varying degrees
by ISM absorption (see, e.g., Figure 3). 
Also listed are \Ha fluxes for each object from Ho et al. (1997a),
Stauffer (1982; NGC 5055), and Keel (1983; NGC 4569).
For NGC 404, NGC 4569, and NGC 4579, we 
have verified the photometric accuracy of 
the nuclear \Ha flux using  narrow-band \hst
WFPC2 images, to be described elsewhere.
The fluxes listed in Table 3 have not been corrected
for Galactic or internal extinction, but in our 
analysis we use fluxes corrected for Galactic extinction
according to the values listed in Table 1.

The conclusion that the UV continua and emission lines are 
uncorrelated is initially
surprising, since it was noted in Maoz et al. (1995)
 that the \Ha fluxes of the UV-bright LINERs in their sample
are well correlated with the 2200~\AA\ continuum, and that
the extrapolation of the observed continuum flux (assuming a
power law with $\beta=-1$) could provide
sufficient ionizing photons to drive the observed \Ha flux. To
reexamine this, we plot in
Figure 5 the \Ha flux  
vs. $f_{\lambda}(1300~{\rm\AA})$
measured from the spectra and corrected for Galactic
extinction.
There appears to be little relation
between these observables. 
This departure from the results of Maoz et al. (1995)
comes about in several ways.
The FOS UV fluxes of NGC~404, NGC~4569, and NGC~5055 
are consistent with those measured with the FOC. 
As discussed by Barth et al. (1996b) and above,
the FOS measurement of NGC 4579  is lower 
by a factor of 3 than the FOC measurement. One may argue that the
NGC~4579 point
in Figure 5 should be plotted at the FOC position (i.e., 
$\log f_{\lambda}(1300~{\rm\AA})=-14.4$, assuming the
$f_{\lambda}(1300~{\rm\AA})
/f_{\lambda}(2200~{\rm\AA})$ ratio stays constant, and correcting for
Galactic extinction), either because the 
FOS measurement is less robust, or because the high flux is the
``normal'' one which the recombining gas ``remembers.''
NGC 4594, M81, and NGC 6500 were not in the Maoz et al. (1995) FOC survey,
and it is they that ruin the correlation that was suggested
by Maoz et al. (1995).
As we will show below, all three of these LINERs
suffer from severe ionizing photon deficits (as already noted
by Ho et al. 1996 for M81 and by Nicholson et al. 1998 for 
NGC 4594). Note that they would {\it not} have
been missed in the FOC
survey and classified as ``UV-dark''; M87 and NGC 4736 (which
were in the FOC survey, but are not in the present sample) have comparable
UV brightnesses, and were easily detected by the FOC (see Maoz 1996).

The FOS measurements provide us with new input for
estimating the ionizing photon budget in these LINERs.
First, we can now measure the UV flux at a wavelength
closer to the Lyman limit. Second, we can measure the UV continuum
slope in the \hst range directly, instead of assuming a power law with $\beta=-1$.
And third, in at least three of the objects, there are unambiguous
spectral signatures of a hot stellar population, so we can
estimate the ionizing flux based on the expectations from such
a population, rather than assume an extrapolated power-law continuum.
As it happens, all three constraints work toward lowering
the estimate of the number of ionizing photons, possibly bringing
into the photon deficit regime 
even those objects for which Maoz et al. (1995) concluded
there was no shortage.
The FOS 2200 \AA\ fluxes measured for NGC 404, NGC 4579, and NGC 5055
are lower than their estimated FOC fluxes (see Table 1).
Except for NGC 4579, all the LINERs have power-law continua
that are softer than $\beta=-1$, leading to a lower 1300 \AA\ or
912 \AA\ flux. For those LINERs whose continua are dominated
by hot stars, there will be fewer ionizing photons
than from a power law  extrapolated beyond the Lyman
limit, because hot stars have a substantial
absorption at the Lyman edge.

We have computed the ratio of the H$\alpha$ line flux to the
the 1300~\AA\ continuum flux density for young star clusters
containing populations of O-type stars.  We carried out our population
synthesis calculations using the Geneva stellar evolutionary tracks
for stars with solar metallicity (Schaerer et al. 1993).  In our models
we employ non-LTE model atmospheres for hot ($T\ge 25,000 K$) stars
(Pauldrach et al. 1994; 1997) to compute the cluster
Lyman continuum fluxes.  The 1300~\AA~ fluxes are computed with the
assumption that the stars radiate as blackbodies at wavelengths 
longward of the Lyman limit.
Further details of these and related computations are described
Tacconi-Garman, Sternberg, \& Eckart (1996) and
Sternberg (1998). Here we consider young clusters
($\ltorder 10^6$ yr old) with Salpeter initial-mass functions (IMFs).
Assuming case-B recombination in $10^4$ K ionization-bounded nebulae, the 
H$\alpha$ line luminosities per number of O-type stars are equal
to 1.4$\times 10^{36}$ and 1.4$\times 10^{37}$ erg s$^{-1}$ for
IMFs which extend up to 30 $M_\odot$ and 120 $M_\odot$, respectively.
For such clusters the H$\alpha/f_{\lambda}(1300$~\AA) ratios equal about 10.5 and 42.0 \AA,
and decrease with increasing cluster age.

 The stellar-wind signatures in several of the
LINERs require the presence of stars of mass $\gtorder 30 M_{\odot}$.
 The two solid diagonal lines in Figure 5 show the maximum \Ha flux
that can be produced with 100\% covering
factor from
 ionization by a stellar population with the given 1300 \AA\ flux
and upper mass cutoff of $120 M_{\odot}$ or $30 M_{\odot}$. 
The two dotted lines show this limit for ionization by power-law continua
of the form $\lambda^{\beta}$ with $\beta=-1$ or $\beta=0$.

 We see that, if there is no internal extinction, then some
and possibly all of the LINERs
require an additional ionizing source to drive their line flux.
M81, NGC 4594, and NGC 6500, even if their UV spectra are extrapolated as power
laws with $\beta=-1$ (i.e., harder than observed) rather than as 
stellar-population spectra, also have a severe ionizing photon
deficit. NGC 4579, on the other hand, does have a $\beta\approx -1$ slope,
as was assumed by Maoz et al. (1995). If the Maoz et al. (1995) flux
level is adopted, then as before, it has a factor 2.8 ionizing
photon surplus, rather than a deficit. If, however, stars dominate
the UV continuum or the FOS flux level is the ``normal'' level,
then this LINER also has an ionizing photon deficit. In the three LINERs
whose UV emission is clearly dominated by stars (NGC 404, NGC 4569,
and NGC 5055), ionization by the stellar population
can provide the required power, but only if very massive stars
are still present.
 Interestingly, it is these three objects which
also have the lowest emission-line ratios
of [O~III]~$\lambda$5007/H$\beta$, [O~I]~$\lambda$ 6300/H$\alpha$, 
and [S~II] $\lambda\lambda$6716, 6731/\Ha in the sample (see Ho et al. 1997a); 
this is exactly what one
would expect from ionization by the relatively soft continuum of a
significant stellar component, which produces less heating per ionization
and a smaller partially ionized zone than an AGN-like power-law
continuum. Stellar photoionization
models for LINERs have been presented by Filippenko \& Terlevich (1992)
and Shields (1992).

Alternatively, some internal foreground extinction (such that
the UV emission is attenuated as seen by the observer, but not 
as seen by the ionized gas) could, in principle, alleviate
the ionization budget problem in some of the objects.
As discussed in \S 3.2, above, the uncertainty in the extinction
curve and the many possible configurations for different extinction
of the continuum and nebular emission result in a broad
range of possible extinction corrections.
To illustrate the effect of a plausible extinction correction, 
Figure 5 shows an  $A_V=0.5$ mag foreground
extinction vector, assuming a Galactic
extinction curve. A Galactic curve is intermediate in ``greyness''
to the SMC and Calzetti et al. (1994) curves considered in \S 3.2.
Note that the objects with the most severe ionizing photon
deficits (M81, NGC 4594, and NGC 6500) 
are those whose Balmer decrements (see Table 3) indicate
little internal reddening, $A_V=0.1$ to 0.35 mag, for the range
in extinction curves.
In NGC 4569, NGC 5055, and
NGC 4579, as discussed in \S 3.2, a plausible extinction correction
will increase in the intrinsic
 1300~\AA\ flux by a factor $\gtorder 10$ (with
 a corresponding increase in the number
of O-type stars),
and may relieve the need for the most massive
 ($\sim 120 M_{\odot}$) stars.

 We conclude that at least some
of the LINERs in our sample have an ionizing photon
deficit, indicating an additional energy source, beyond
that implied by the observable UV.
To search for evidence for such an additional 
source, we have compiled X-ray data for the seven LINERs.
For NGC 404, NGC 4569, NGC 4579, and NGC 4594 we have
used analysis of both archival and new {\it ASCA} data by
Y. Terashima, A. Ptak, and L. Ho (to be described elsewhere) to
obtain unabsorbed 2--10 keV fluxes. For M81 we have used the
{\it ASCA} flux derived by Ishisaki et al. (1996). 
NGC 5055 and NGC 6500 have not been observed with {\it ASCA},
but have been observed by {\it ROSAT} in the 0.1--2.4 keV
band. We use the {\it ROSAT} HRI flux
derived by Barth et al. (1997) for NGC 6500 and 
extrapolate
it to the {\it ASCA} bandpass,
assuming an X-ray flux density $f_{\nu}\propto\nu^{-0.7}$. 
Read, Ponman, \& Strickland (1997) have observed NGC 5055 with the
 {\it ROSAT} PSPC, 
and find that the nuclear source is best fit with a 
cool (0.74 keV) thermal plasma, absorbed by a hydrogen column 
of $2.55\times 10^{20}$ cm$^{-2}$, and a flux
of $3.3 \times 10^{-13}$ erg s$^{-1}$ cm$^{-2}$ escaping
from the galaxy. We extrapolate this model to find
the unabsorbed 2--10 keV band flux. We caution that the
extrapolated fluxes for these two galaxies are uncertain.
In the case of NGC 6500, both the instrument (HRI) and the low
count statistics preclude any spectral information. In
NGC 5055, it is possible that there is a power-law component 
which dominates over the thermal component in the 2--10 keV band.
Table 3 lists our assumed 2--10 keV fluxes for the seven LINERs.

Figure 6 shows the ratio of X-ray-to-UV power vs. 
H$\alpha$-to-UV power for each of the LINERs. Note how in the
four ``UV-photon starved'' objects (high \Ha to UV power
ratio) the X-ray power is comparable to or greater than
the UV power. Conversely, in the three objects without
a serious ionizing photon deficit, the X-ray power is one to two
orders of magnitude lower than the UV power.
This suggests that, indeed, an energy source emitting
primarily in the extreme-UV, and not directly related to the
observed UV source, may be the main ionizing agent in the
four ``UV-photon starved'' objects. The observed X-ray emission
would be the high-energy tail of such a component.
For example, a blackbody with temperature $>3.3\times 10^5 {\rm K}$ 
would have
an X-ray to UV power ratio greater than or equal to that of the
UV-photon starved objects. Alternatively, the photoionizing continua in
 these objects could consist
of power-law spectra extending from the X-rays to the Lyman limit.
However, this would require that the central UV sources 
are significantly attenuated by dust extinction while the nebular 
components are not.

\subsection{The Ionization Mechanism}

It has long been debated whether the emission lines in
LINERs are produced by means of photoionization or shocks.
The question has been recently re-addressed in the analysis
of the four LINERs among the seven discussed here with
published UV spectra (Ho et al. 1996; Barth et al. 1996b, 1997;
Nicholson et al. 1998). These studies show that the UV line
ratios are consistent with either photoionization by an
AGN-like continuum or by slow-moving shocks, but inconsistent
with the fast ``photoionizing'' shocks proposed by Dopita \& Sutherland (1996).

The three additional LINERs introduced in the present study
 have either very weak (NGC 404) or 
undetected (NGC 4569, NGC 5055) UV line emission. 
For this reason, we choose not to carry out detailed
modeling of these three objects. In principle, one
could study the ratios, or the lower limits of the ratios,
between optical and UV emission lines. However, such
ratios are extremely sensitive to the assumed extinction
corrections, especially in NGC 4569 and NGC 5055, which
have moderately steep Balmer decrements (see Table 3 and discussion 
in \S 3.2). The uncertainty in the magnitude and form
of the extinction makes it difficult to exclude any
models based on optical/UV line ratios.

We note that the ratios of the weak UV lines
measured in NGC 404 are similar to those of the previously studied
LINERs, so the conclusion regarding the viability of 
photoionization and slow-moving shocks holds for this object as well. 
In NGC 4569 and NGC 5055,
the complete absence of UV emission lines argues against
shock excitation, since a rich UV emission-line spectrum
would be expected, given the observed optical line spectrum.
Furthermore, the absorption lines in the spectra of these objects
show that massive stars exist in the nuclei. The sources
required to produce most or all of the emission lines by
photoionization are therefore present.  

\section{Conclusions}

We have studied the \hst UV spectra of seven LINERs having
compact nuclear UV sources. Our main findings are as follows.\\
1. At least three of the LINERs have clear spectral signatures
indicating that the dominant UV continuum source is a cluster of massive stars.\\
2. A similar continuum source could dominate in the other four
LINERs as well, but its spectral signatures would be veiled by
low S/N or superposed broad emission lines. 
Alternatively, there may be two types of UV-bright LINERs: those 
where the UV continuum is produced by a starburst, and those where
it is nonstellar. If the variability
of NGC 4579 is real, its continuum source is certainly an AGN.\\
3. The seven LINERs have similar UV SEDs, which are approximately
flat in $f_{\lambda}$. The equivalent widths of the UV emission lines
span two orders of magnitude. In the objects with the highest S/N,
the continuum shape, the UV interstellar absorption lines, the optical line ratios,
and the X-ray absorbing column are all broadly consistent with 
a starburst spectrum extinguished by $A_V\approx 0.5$ mag.\\
4. The three ``clearly-stellar'' LINERs
have relatively weak X-ray emission, and their stellar populations probably
provide enough ionizing photons to explain the observed optical
emission-line flux. The four other LINERs have severe ionizing
photon deficits, for reasonable extrapolations of
 their UV spectra beyond the Lyman limit, but an X-ray/UV power ratio
that is higher by two orders of magnitudes than that of the
three stellar LINERs. A component
which emits primarily in the extreme-UV may be the main 
photoionizing agent in these four objects. 

The picture emerging from this comparison is that the compact 
UV continuum source seen in $\sim 25\%$ of LINERs (Maoz et al. 1995;
Barth et al. 1996a, 1998) is, at least in some cases, a nuclear 
starburst rather than an AGN-like nonstellar object. The UV luminosity
is driven by tens to thousands of O-type stars, depending on
the object and the extinction assumed. The O-stars could be the high-mass
end of a bound stellar population, similar to those seen in
super star clusters (e.g., Maoz et al. 1996b) .
 Nonstellar
sources in LINERs may be significant or even dominate at other 
wavelengths, as we indeed find for some of the objects. Even
the three ``clearly-stellar'' LINERs, which do not obviously require
the existence of such an additional source, may well have one; the
ionizing photon budget estimate was made assuming a 100\% covering
factor of the line-emitting gas, which is not necessarily true.
This picture fits well with recent results showing that nuclear-starburst
and quasar-like activity are often intermingled in Seyfert 1 and 2
galaxies. Our results extend this result to the lower luminosities
of the LINERs discussed here, although the question of whether a 
``micro-quasar'' exists in these objects is still open. Apparently,
the UV is not the best place to look for micro-quasars in most LINERs.

Higher S/N UV spectra of some of the objects in our sample could reveal
whether they too have the signatures of massive stars. Conversely,
further evidence of UV continuum variability, as suggested in NGC 4579,
should be sought in LINERs. Even in those LINERs with stellar signatures,
UV variability could reveal a contribution by a nonstellar component.
Our work suggests, however, that the AGNs possibly associated with LINERs 
are most prominent at higher energies.
X-ray observations by upcoming space missions, having better angular and spectral
resolution and higher sensitivity, will likely provide key insights
to the nature of LINERs.

\acknowledgements
We thank A. Ptak and Y. Terashima for generously providing {\it ASCA}
measurements and analysis prior to publication.
M. Eracleous and R. Pogge are thanked for permission
to use \hst WFPC2 imaging data and analysis prior to
publication. We are grateful to D. Calzetti, T.M. Heckman, A.J. Barth,
and an anonymous referee for
valuable input.
This work was supported by grants GO-6112 and AR-5712
 from the Space Telescope
Science Institute, which is operated by AURA, Inc., under NASA
contract NAS 5-26555. NASA grant NAG 5-3556 is also acknowledged.
D. M. is grateful for the hospitality of the STScI Visitor Program
in the course of this
work. D.M. and A.S. acknowledge support by  the U.S.-Israel
Binational Science Foundation grant 94-00300 and by the Israel
Science Foundation.

\appendix

\section{Interstellar Absorption Lines in NGC 4569}

The high S/N of the NGC 4569 spectrum reveals numerous
interstellar absorption lines.
Table 2 lists the strongest detected lines.
The spectrum shows narrow absorption features of C~I
$\lambda\lambda$1261, 1277, 1280, 1329, 1561, and 1657 \AA,  consistent with
absorption at the velocity of the galaxy.  The
equivalent widths of these lines range from 0.5 \AA\ for $\lambda$1280
to 3.5 \AA\ for $\lambda$1657. C~I  absorption lines in the UV are often present in the
spectra of starbursts, but with equivalent widths that are
usually less than 1 \AA\ (T. Heckman, private communication).
Absorption by C~I is noteworthy since its ionization potential 
is only 11.26 eV; survival of neutral carbon
thus requires an environment with substantial shielding or dilution of
the radiation at wavelengths longward of the Lyman
limit.

UV absorption lines of C~I  have been observed in
Galactic diffuse clouds along several
lines of sight (e.g., Joseph et al. 1986).
 The atomic
carbon abundance becomes even larger in thicker translucent clouds and 
at the edges of dense
photon-dominated regions which
are hard to observe in UV absorption (due to their high optical depths).  Recent work
on C~I detections in our Galaxy and in starburst galaxies
 has focussed on the two submillimeter fine-structure transitions emitted
by the ground-state triplet at 492.2 and 809.3 GHz (e.g., Stark \& van Dishoeck 1994;
Stutzki et al. 1997).

The measured absorption lines can be used to place a bound on the
C~I column density in NGC~4569.  If we assume that absorption occurs
on the linear part of the curve of growth, the resulting values
for the column density are
inversely correlated with oscillator strength for the six measured
transitions, indicating at least partial saturation in the lines.  We
can then place a conservative lower limit on the column density
$N({\rm C~I})$ from the transition with the weakest oscillator
strength, $\lambda$1280, which implies $N({\rm C~I}) \simgt 2 \times
10^{15}$ cm$^{-2}$.  For solar abundances modified by a representative
interstellar depletion factor of 0.4 dex (Savage \& Sembach 1996),
this result implies an accompanying column density of hydrogen $N({\rm
H}) \simgt 10^{19}$ cm$^{-2}$, divided by the carbon
neutral fraction.

Similar absorption is also seen in Fe~II transitions at
$\lambda\lambda$1608, 2344, 2374, 2383, 2587, and 2600 \AA, with
equivalent widths of $\sim 1 - 3$ \AA.  These lines also show
evidence for saturation, and can provide a lower limit on $N({\rm
Fe~II})$. The
velocity structure of the lines indicates a Galactic contribution
of roughly a quarter of the
total absorption; removing this contribution yields $N({\rm Fe~II})
\simgt 6 \times 10^{14}$ cm$^{-2}$.  For solar abundances, the
corresponding hydrogen bound is $N({\rm H}) \simgt 2 \times 10^{19}$
cm$^{-2}$, but this limit would increase to $N({\rm H}) \simgt
10^{21}$ cm$^{-2}$ for depletion similar to that in the local
ISM.  Significant absorption ($\simgt 0.5$ \AA) is
{\sl not} detected in the accessible Fe~I features, implying $N({\rm
Fe~I}) \simlt 2 \times 10^{13}$ cm$^{-2}$ and an Fe~II/Fe~I ionic
ratio $\simgt 30$.

The H$\alpha$ emission-line flux observed in NGC 4569  suggests that the
covering factor of emission-line clouds is high
(cf. Fig. 5), and the possibility thus arises that the medium
responsible for absorption by C~I, Fe~II, and other species (Fig. 2)
can be identified with such clouds along our line of
sight to the continuum source(s).  A coherent picture along these
lines appears to be possible, with the absorption in NGC~4569
resulting from material with a total column density $N({\rm H~I +
H~II})$ of a few times $10^{21}$ cm$^{-2}$.  Standard recombination
arguments for ionization-bounded clouds imply that the column density
of ionized hydrogen $N({\rm H~II})$, expressed in cm$^{-2}$, scales
approximately as log $N({\rm H~II})$ = log $U$ + 23, where $U$ is the
ionization parameter (defined here as the dimensionless ratio of
ionizing photon density to particle density at the irradiated cloud
face; see Voit 1992).  For LINERs, log $U \approx -3$ to $-4$ (e.g.,
Ferland \& Netzer 1983), so that $N({\rm H~II}) \approx 10^{19} -
10^{20}$ cm$^{-2}$.  This zone will thus constitute an ionized surface
on one side of the cloud, with the bulk of the cloud comprised of
neutral or weakly-ionized material.  The C~I and Fe~II column
densities as well as the Fe~II/Fe~I ratio will depend on the detailed
continuum shape as well as dust shielding, but tests with the
photoionization code CLOUDY (Ferland 1996) suggest that the measured
bounds are easily plausible in this model, assuming solar abundances
and standard depletion factors.

An absorbing column density of this type should have measurable
effects on the continuum.  Dust associated with the absorbing gas
will produce extinction of order $A_V \approx
1$ mag, as is indeed suggested in this galaxy (see \S 3.2).
 The absorbing medium
will also imprint bound-free absorption on the soft
X-ray emission from NGC 4569.  Preliminary fits to {\it ASCA}
spectra of this object suggest that substantial absorption is, in
fact, present, with a total hydrogen column density
(assuming solar abundances) of order $10^{22}$ cm$^{-2}$ 
(Terashima 1998), in
plausible agreement with this picture.

\begin{deluxetable}{crllrrrrrrr}
\scriptsize
%\tiny
\tablewidth{0pt}
\tablecaption{Journal of Observations}
\tablehead{\colhead {Galaxy} & \colhead {$v_h$} & 
\colhead {UT Date} & \colhead {Aperture} & \multicolumn{3}{c}
{Exp. Time (s)}&\multicolumn{3}{c}{Observed $f_{\lambda}(2200 {\rm \AA})^a$}&\colhead{Galactic} \nl
\colhead {}&\colhead {km s$^{-1}$}&\colhead {}&\colhead {}&\colhead {G130H}&
\colhead {G190H}&\colhead {G270H}
&\colhead{FOS}&\colhead{FOC$^b$}&\colhead{WFPC2$^c$}&\colhead{$E(B-V)$}
\nl}
\startdata
NGC 404 & $-36$&1994 Dec 25   &$0.86''$&6400&2130&540     & 115    &180    &\nodata&0.055$^d$\nl
M81     & $-28$&1993 Apr 5,14 &$0.30''$&3000&\nodata&2000 & 150$^e$&\nodata&200$^e$&0.035$^d$\nl
NGC 4569&$-223$&1996 Apr 11   &$0.86''$&4790&1320&690     &1050    &1000   &1100   &0.050$^f$\nl
NGC 4579& 1500&1994 Dec 16   &$0.43''$&6840&3300&690     &33      & 110   &\nodata&0.052$^f$\nl
NGC 4594& 1128&1995 Jan 12   &$0.86''$&3720&1800&480     &12      &\nodata&\nodata&0.066$^f$\nl
NGC 5055&  497&1996 Apr 12   &$0.86''$&7380&2510&2440    &77      &100    &\nodata&0.000$^d$\nl
NGC 6500& 3003&1994 Aug 13   &$0.86''$&10260&3110&1020   &27      &\nodata&27     &0.093$^d$\nl
\enddata							     
\tablecomments{ \\
$^a$ In units of $10^{-17}$ erg s$^{-1}$ cm$^{-2}$ \AA$^{-1}$.\\
$^b$ From Maoz et al. (1995).\\
$^c$ From Barth et al. (1998).\\
$^d$ From Burstein \& Heiles (1984).\\
$^e$ At 1500 \AA, rather than 2200 \AA, and computed
from the WFPC2 counts by assuming a scaled
version of the 1600--2300~\AA\ spectrum of NGC 4579.\\
$^f$ From Murphy et al. (1996).}
\end{deluxetable}						     

\begin{deluxetable}{rllll}
\scriptsize
%\tiny
\tablewidth{300pt}
\tablecaption{Absorption Lines}
\tablehead{\colhead {Line} & 
\multicolumn{4}{c}{$W_{\lambda}^a$}
\nl
\colhead{}&\colhead{NGC 4569}&\colhead{NGC 404}&\colhead{NGC 5055}
&\colhead{Milky Way}\nl}
\startdata
N~V   $\lambda$1240$^b$ &      2.8 &3.5&\nodata&\nodata     \nl
Si~II $\lambda$1260&      2.6 &1.7 & 1.8 & 0.6-1.4     \nl
C~I $\lambda$1261 &      $1.0^c$ &\nodata&\nodata&\nodata  \nl
C~I $\lambda$1277 &       1.4 &0.7&\nodata&\nodata     \nl
C~I $\lambda$1280 &       0.6 &\nodata&\nodata&\nodata     \nl
Si~III $\lambda\lambda$1294, 1297$^d$&  0.8 &\nodata&\nodata&\nodata     \nl
O~I $\lambda$1302+Si~II $\lambda$1304&     4.2 &\nodata& 4.0& 0.8-1.7     \nl
C~I $\lambda$1329 &       1.2 &\nodata&\nodata&\nodata     \nl
C~II  $\lambda$1335&      3.6 &2.0&\nodata&0.8-1.0     \nl
Si~IV $\lambda\lambda$1394, 1403$^b$ &   8.3&2.9&7.8&0.8      \nl
Si~III $\lambda$1417$^d$&     0.5  &\nodata&\nodata&\nodata    \nl
C~III  $\lambda$1427$^d$&     0.4  &\nodata&\nodata&\nodata    \nl
S~V    $\lambda$1502$^d$&     1.3  &\nodata&\nodata&\nodata    \nl
Si~II  $\lambda$1527&     2.3 &\nodata&1.6&0.5-0.7     \nl
C~IV   $\lambda$1549$^b$ &     11.&4.8&15.&0.3-0.5      \nl
C~I $\lambda$1561 &      $1.2^e$ &0.8&\nodata&\nodata  \nl 
Fe~II $\lambda$1608&     $1.5^f$&\nodata&\nodata&\nodata   \nl       
C~I $\lambda$1657 &       3.6 &\nodata&\nodata&\nodata     \nl
Al~II $\lambda$1670&      3.5&2.5&2.1&0.6-0.8      \nl
N~IV   $\lambda$1720$^d$&     1.6&\nodata&\nodata&\nodata      \nl
Fe~II $\lambda$2344&     $2.9^f$&2.2&2.1&0.7-1.2   \nl       
Fe~II $\lambda$2374&     $2.1^f$&1.2&\nodata&0.6-1.0   \nl  
Fe~II $\lambda$2383&     $3.4^f$&1.4&3.2&0.8-1.4   \nl       
Fe~II $\lambda$2586&     $2.4^f$&1.8&2.0&0.6-1.1   \nl       
Fe~II $\lambda$2600&     $3.3^f$&2.5&2.6&0.9-1.5   \nl       
Mg~II $\lambda$2800&      9.9&7.5&9.6&2.1-3.4      \nl
Mg~I  $\lambda$2852&      1.8&1.7&2.7&0.4-0.6      \nl
\enddata
\tablecomments{
$^a$ Equivalent width, in \AA. Milky Way range, based on quasar
spectra, from Savage et al. (1993).\\
$^b$ Interstellar, plus broad blueshifted absorption from O-type star atmospheres.\\
$^c$ After approximate deblending from Si~II $\lambda$1260.\\
$^d$ Non-resonance line, originating in the photospheres of hot stars.\\
$^e$ Difficult to determine the continuum level because of C~IV $\lambda$1549
absorption.\\
$^f$ From the line profile, roughly 1/4 of the absorption is Galactic.}
\end{deluxetable}						     

\begin{deluxetable}{rrrrrrrr}
\footnotesize
%\tiny
\tablewidth{0pt}
\tablecaption{Observed Emission}
\tablehead{\colhead {Emission} & 
\colhead {} & \multicolumn{6}{c}
{NGC} \nl
\colhead {}&\colhead {M81}&\colhead {4579}&\colhead {6500}&
\colhead {4594}&\colhead {404}
&\colhead{5055}&\colhead{4569}
\nl}
\startdata
C IV    $\lambda 1549$       & 140   & 70.0 &  $<1.0$  & $<1.6$ &$<2.0$&$<2.0$  &$<2.0$ \nl
He II   $\lambda 1640$       &\nodata&  5.8 &    1.9   & $<1.4$ & 3.5  &$<2.0$  &$<2.0$ \nl
C III]  $\lambda 1909$       &\nodata& 75.0 &    4.3   &   2.9  & 1.9  &$<1.0$  &$<1.0$ \nl
N II    $\lambda 2141$       &\nodata&  6.4 &  $<0.4$  & $<0.5$ &$<1.0$&$<1.0$  &$<1.0$ \nl
C II]   $\lambda 2326$       & 180   & 44.0 &    3.6   &   2.8  & 3.5  &$<1.0$  &$<1.0$ \nl
[O II]  $\lambda 2470$       &  31   &  6.5 &    1.1   &   0.7  &$<1.0$&$<1.0$  &$<1.0$ \nl
\Ha $\lambda 6563^a$         & 1150  &  165 &    110   &   105  & 62   &  43    &  440  \nl
\Ha/H$\beta^a$               & 3.46  &  5.26&    3.65  &   3.46 & 3.72 & 5.42   &  5.04 \nl
$f_{\lambda}(1300{\rm\AA})^b$&  15   &    8 &    4     &    3   &  16  &   7    &   75  \nl 
2--10 keV$^c$                & 120   &   42 &$13\pm5^d$&   29   &$<3$  &$0.13^d$&$4\pm2$\nl 
$\beta^e$                    &   0   &$-1.5$&$-0.5$    &\nodata &$-0.5$&   0    &   0.5 \nl
\enddata
\tablecomments{Except for X-ray fluxes, 
values are uncorrected for Galactic or internal extinction.
Line fluxes in units of $10^{-15}$ erg s$^{-1}$ cm$^{-2}$.\\
$^a$ From Ho et al. (1997a), Stauffer (1982), and Keel (1983).\\
$^b$ In units of $10^{-16}$ erg s$^{-1}$ cm$^{-2}$ \AA$^{-1}$.\\
$^c$ In units of $10^{-13}$ erg s$^{-1}$ cm$^{-2}$, after correction
for absorption.\\
$^d$ Extrapolated from {\it ROSAT} 0.1--2.4 keV data; see text.\\
$^e$ Spectral slope, $f_{\lambda}\propto \lambda^{\beta}$, in the range
1250--2300 \AA.}
\end{deluxetable}						     

\begin{figure}

\plotfiddle{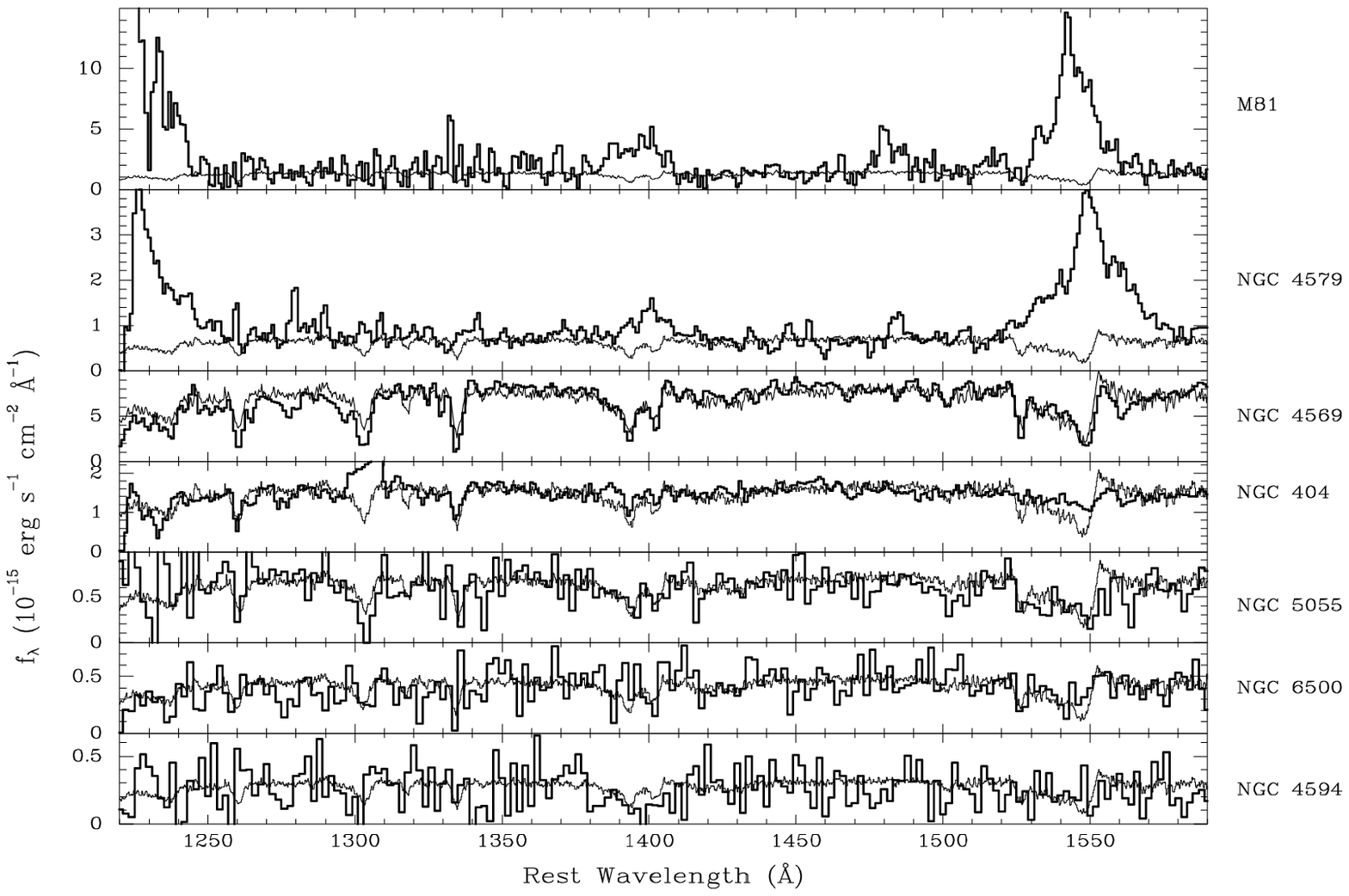}{425pt}{0}{120}{120}{-380}{-200}
\caption{FOS G130H spectra of the seven LINERs (bold lines), ordered with the
two broad-lined objects on top, and then with decreasing $f_{\lambda}$.
Overlayed in each case is the spectrum of the
starburst in NGC 1741-B, normalized to be flat in $f_{\lambda}$ and scaled
by a multiplicative factor to match the LINER continuum level.
}
\end{figure}

\begin{figure}
\epsscale{1.7}
\plotone{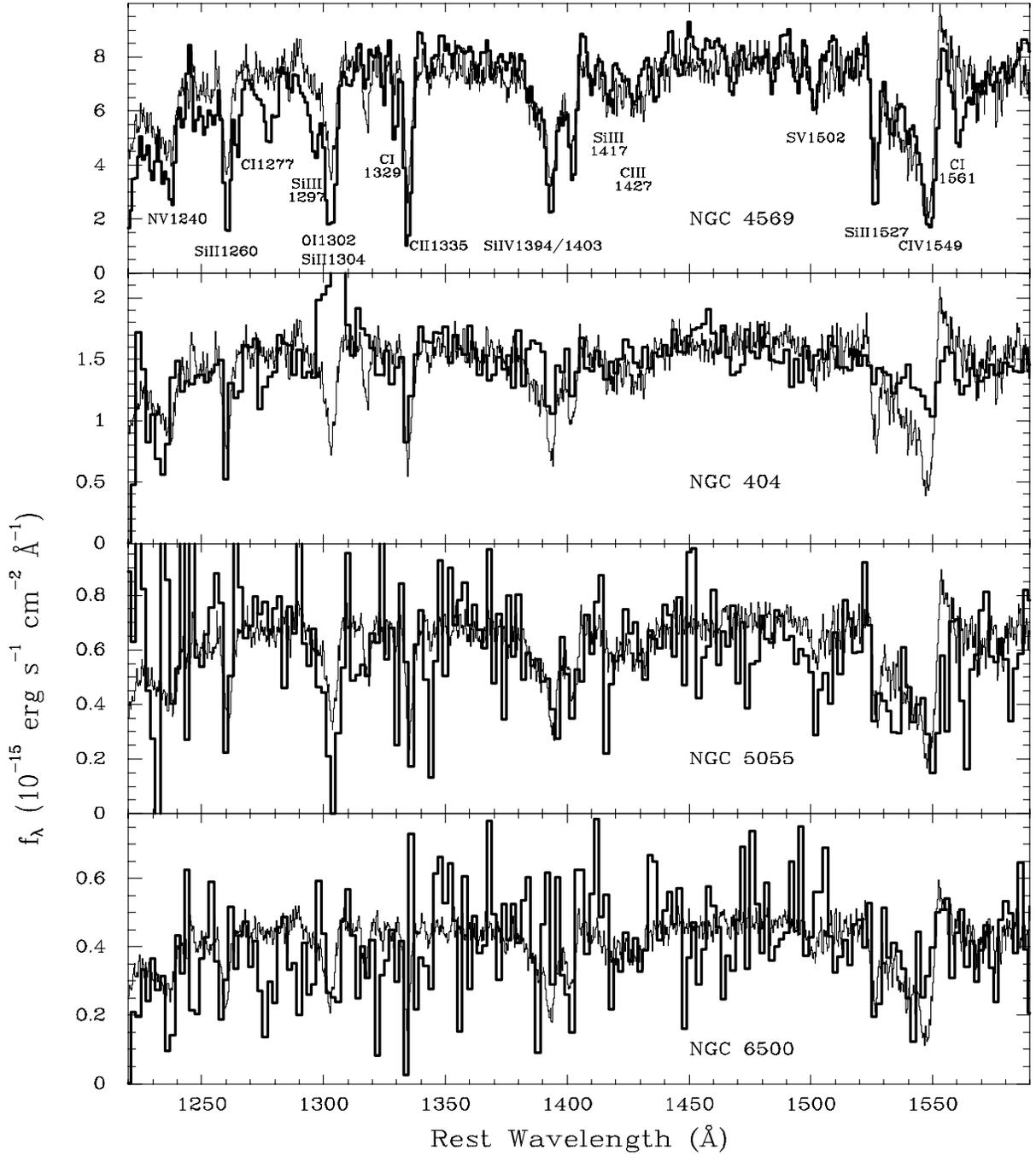}
\caption{Same as Figure 1, but only for the four brightest
LINERs devoid of broad emission lines. The strongest absorption
lines are marked for NGC 4569. The emission feature near 1300~\AA\ in
NGC 404 is an artifact.
}
\end{figure}

\begin{figure}
\epsscale{1.5}
\plotone{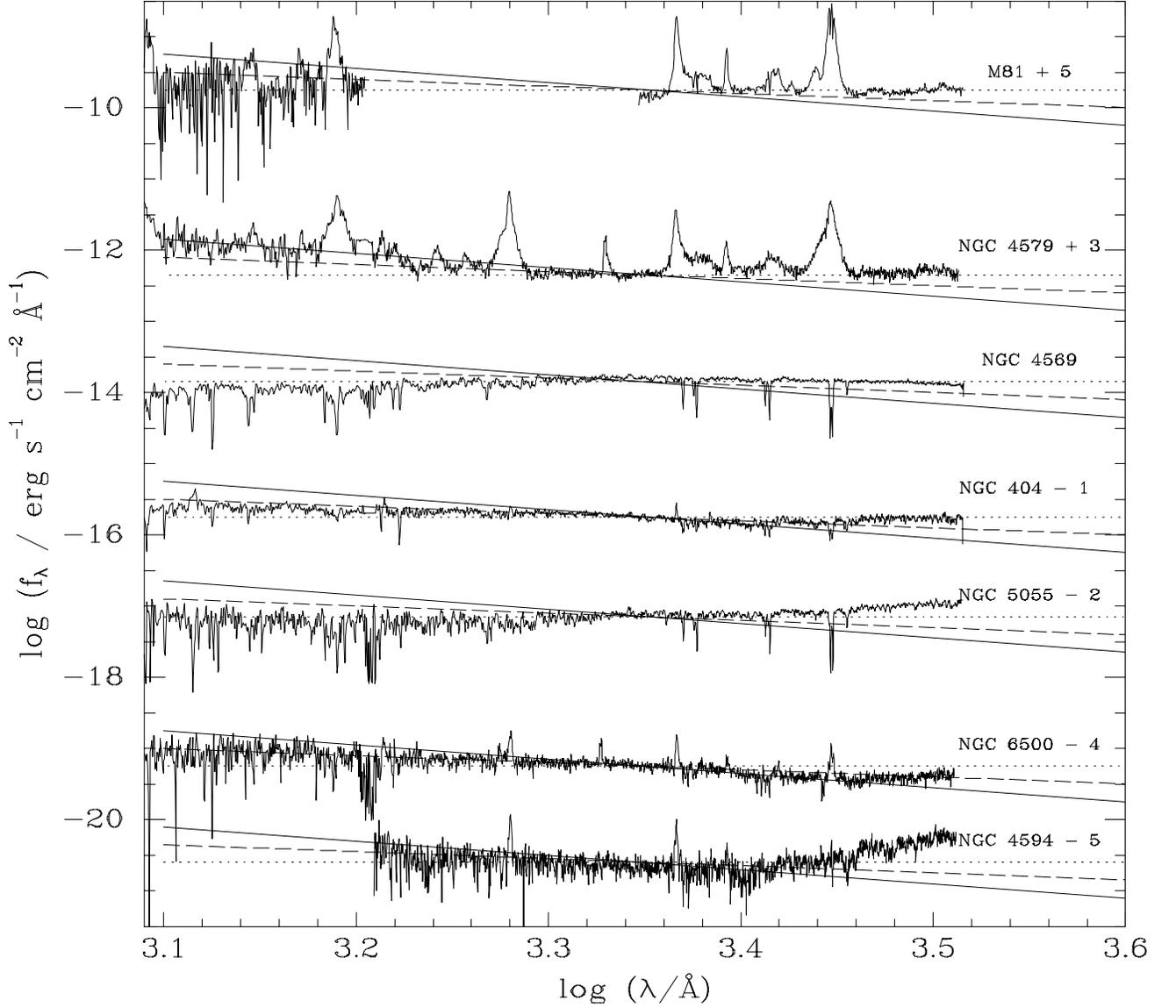}
\caption{UV (1250 -- 3200 \AA) spectral energy distribution (SED) for the
seven LINERs in $\log f_{\lambda}$ vs. $\log \lambda$.
The spectra have been corrected for Galactic extinction
assuming
the $E(B-V)$ values listed in Table 1, and
vertically shifted for clarity, as indicated
on the right side.
The straight lines 
drawn through every spectrum at $\log\lambda=3.36$ (2300 \AA)
are power laws $f_{\lambda}\propto \lambda^{\beta}$, with
$\beta=0,-1, -2$. The seven LINERs
have similar UV spectral slopes. Excluding NGC 4579, they can be 
roughly described shortward of 2300 \AA\ by power laws 
with $\beta \approx 0\pm 0.5$.
}
\end{figure}

\begin{figure}
%\epsscale{1.2}
\plotfiddle{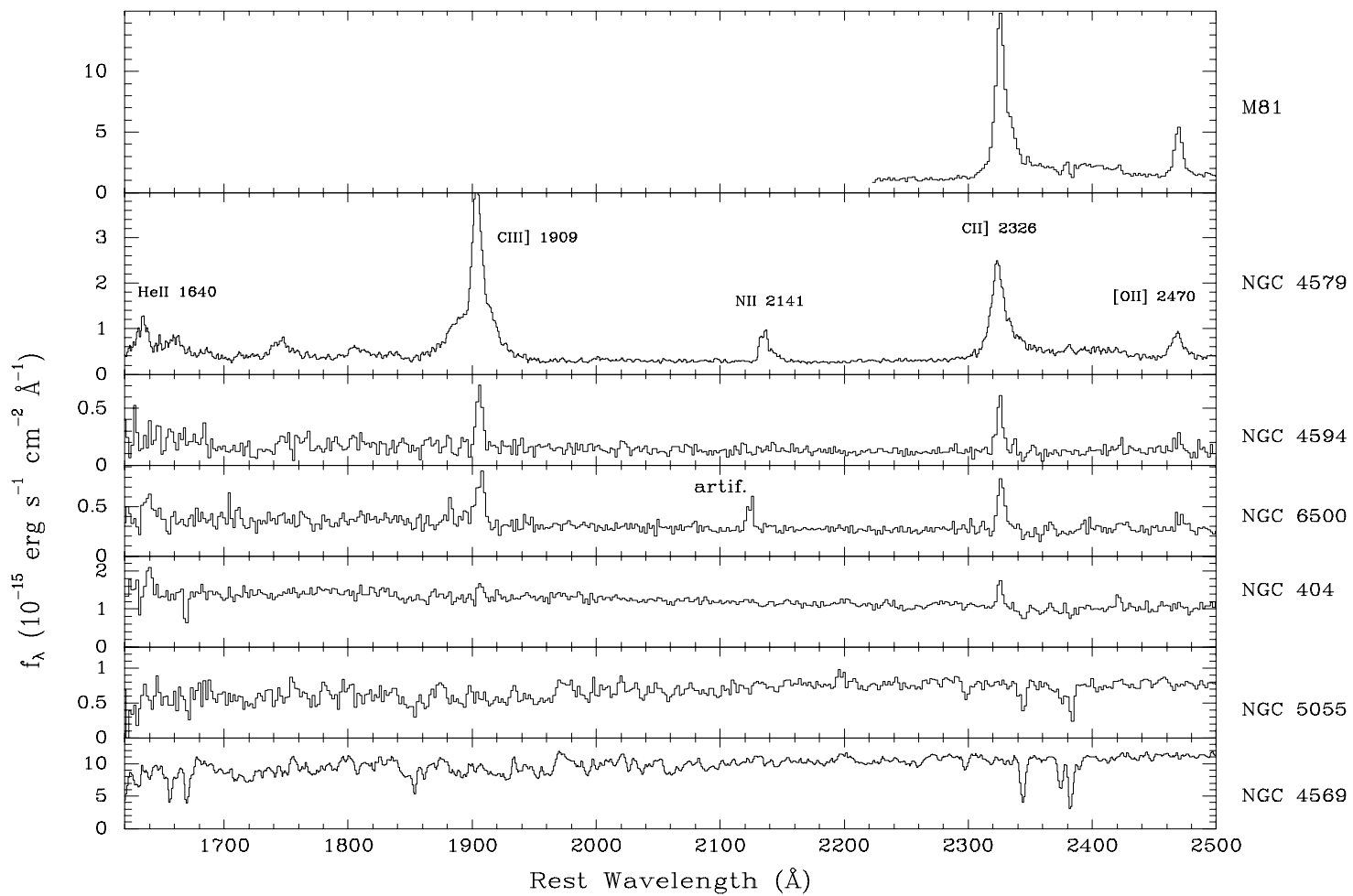}{425pt}{0}{120}{120}{-380}{-200}
\caption{FOS spectra of the seven LINERs in 
the 1620 -- 2500 \AA\ range. The objects are
ordered from top to bottom with decreasing equivalent
width of the main lines, which
are marked in the spectrum of NGC 4579.
}
\end{figure}

\begin{figure}
\epsscale{1.1}
\plotone{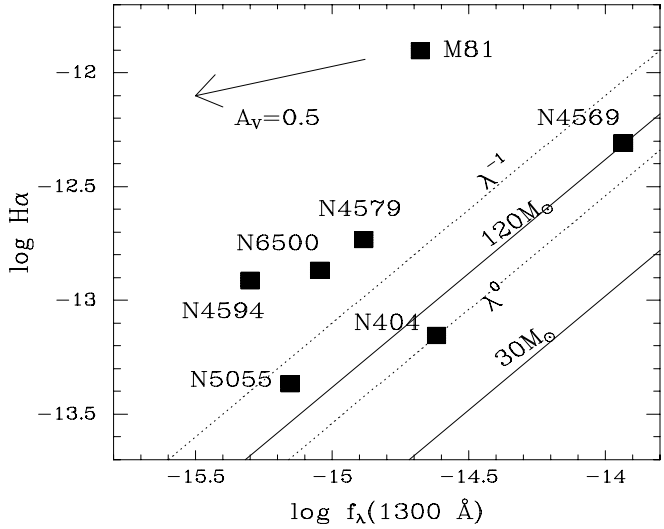}
\caption{Log \Ha flux (in 
erg s$^{-1}$ cm$^{-2}$)  
vs. log $f_{\lambda}(1300{\rm \AA})$ (in 
erg s$^{-1}$ cm$^{-2}$ \AA$^{-1}$)
measured from the spectra and corrected for Galactic
extinction.
The two solid diagonal lines show the maximum \Ha flux
that can be produced in Case B recombination with 100\% covering
factor from
 ionization by a stellar population with a given 1300 \AA\ flux,
resulting from an instantaneous burst of 
 age 1 Myr, Salpeter initial-mass
function, and upper mass cutoff of $120 M_{\odot}$ or $30 M_{\odot}$. 
The two dotted lines show this limit for ionization by power-law continua
$f_{\lambda}\propto\lambda^{\beta}$ with $\beta=-1$ or $\beta=0$.
An  $A_V=0.5$ mag foreground extinction vector, assuming a Galactic
extinction curve, is shown for reference. Typical uncertainties
are 10\% in $f_{\lambda}(1300~{\rm\AA})$ ($\pm 0.04$ in the log) and 
30\% in \Ha flux ($\sim \pm 0.13$ in the log).}
\end{figure}

\begin{figure}
\epsscale{1.1}
\plotone{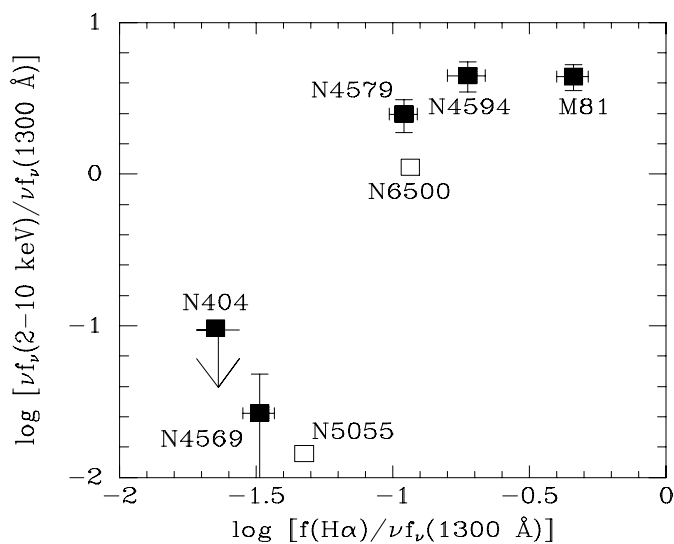}
\caption{The ratio of X-ray-to-UV power vs. 
H$\alpha$-to-UV power for each of the LINERs.
Filled symbols denote {\it ASCA} measurements, empty
symbols are extrapolated {\it ROSAT} values.
 Note how in the
four ``UV-photon starved'' objects (high H$\alpha$-to-UV power
ratio) the X-ray power is comparable to, or greater than,
the UV power. An energy source emitting
primarily in the X-rays, and not necessarily related to the
observed UV source, is probably the main ionizing agent in the
four ``UV-photon starved'' objects.
}
\end{figure}

\end{document}